\documentclass{llncs}

\usepackage{mathptmx}
\usepackage{helvet}
\usepackage{graphicx}
\usepackage{fullpage}
\usepackage{url}
\usepackage{listings}
\usepackage{amssymb}
\usepackage{paralist}
\usepackage{hyperref}

\newcommand{\xf}[1]{Figure~\ref{#1}}

\newcommand{\xs}[1]{Section~\ref{#1}}
\newcommand{\xa}[1]{Appendix~\ref{#1}}

\newcommand{\xt}[1]{Table~\ref{#1}}
\newcommand{\xl}[1]{Listing~\ref{#1}}

\newcommand{\gee}{{GEE\index{GEE}\index{Frameworks!GEE}}}

\newcommand{\gipsy}{{GIPSY\index{GIPSY}}}

\newcommand{\gipl}{{GIPL\index{GIPL}}}
\newcommand{\sipl}{{SIPL\index{SIPL}}}

\newcommand{\lucid}{{Lucid\index{Lucid}}}
\newcommand{\ilucid}{{Indexical Lucid\index{Indexical Lucid}}}

\newcommand{\olucid}{{Objective Lucid\index{Tensor Lucid}}}

\newcommand{\flucid}{{Forensic Lucid\index{Forensic Lucid}}}

\newcommand{\lucx}{{Lucx\index{Lucx}}}

\newcommand{\jooip}{{JOOIP\index{JOOIP}}}
\newcommand{\marfl}{{MARFL\index{MARFL}}}

\newcommand{\olucidop}[1]{{\bf \texttt{\textmd{\textsc{#1}}}}}
\newcommand{\lucidop}[1]{{\bf \texttt{#1}}}
\newcommand{\molucidop}[1]{\mathrm{\;}{\bf \texttt{\textmd{\textsc{#1}}}}\mathrm{\;}}

\newcommand{\trans}{$\psi$}

\newcommand{\invtrans}{$\Psi^{-1}$}

\newcommand{\tab}[1]{\hspace{#1pt}}

\newcommand{\shrule}[0]{\vspace{3pt}\hrule\vspace{6pt}}
\newcommand{\ehrule}[0]{\vspace{6pt}\hrule\vspace{3pt}}

\newcommand{\file}[1]{\url{#1}\index{Files!#1}}

\newcommand{\api}[1]{\texttt{#1}\index{API!#1}}

\newcommand{\statement}[2]
{
	\vspace{7pt}
	\shrule
	#1
	\ehrule
	\vspace{7pt}
}

\newcommand{\sproposition}[1]
{
	\statement{\begin{proposition}#1\end{proposition}}
}

\newcommand{\sdefinition}[1]
{
	\statement{\begin{defn}#1\end{defn}}
}

\newcommand{\saxiom}[1]
{
	\statement{\begin{axioms}#1\end{axioms}}
}

\newcommand{\slemma}[1]
{
	\statement{\begin{lemma}#1\end{lemma}}
}

\newtheorem{defn}{Definition}
\newtheorem{axioms}{Axiom}

\newcommand{\lucidL}[1]{{$\mathit{Lucid}$}($L$) }

		{}

\def\myvert{\raise 2.27pt \hbox{\vrule depth 0pt height 8pt width 0.2mm}}
\def\myarrow{\hspace*{0.43mm}%
             \raise 2.29pt\hbox{\vrule depth 0pt height 8pt width 0.16mm}%
             \hspace*{-0.32mm}%
             $\longrightarrow$
             \ %
             }

\newcommand{\johndef}{\mathcal{D}}

\newcommand{\myid}{\textit{id}}

\newcommand{\mydagger}{\!\dagger\!}
\newcommand{\context}[2]{\mathcal{D},\mathcal{P} \vdash #1 : #2}

\newcommand{\qcontext}[2]{\mathcal{D},\mathcal{P} \vdash #1 \::\: #2}

\newcommand{\myifthenelse}{\mathtt{if}\;E\;\mathtt{then}\;E'\;\mathtt{else}\;E''}

\def\Lat{\index{a@{\texttt{\char64}}}\;\texttt{\char64}\;}
\def\LSat{\index{a@{\texttt{\char64}}}\texttt{\char64}}
\def\Lhash{\index{a@{\texttt{\char35}}}\texttt{\char35}}

\def\Lif{\index{ifthenelse@{\texttt{if then else}}}\texttt{if}\;}
\def\Lthen{\;\texttt{then}\;}
\def\Lelse{\;\texttt{else}\;}

\def\mfirst{\index{first@{\texttt{first}}}\mathrm{{\mathtt{first}}}\;}
\def\mprev{\index{prev@{\texttt{prev}}}\mathrm{{\mathtt{prev}}}\;}
\def\mnext{\index{next@{\texttt{next}}}\mathrm{{\mathtt{next}}}\;}
\def\mfby{\index{fby@{\texttt{fby}}}\;\mathrm{{\mathtt{fby}}}\;}
\def\mwvr{\index{wvr@{\texttt{wvr}}}\;\mathrm{{\mathtt{wvr}}}\;}
\def\mupon{\index{upon@{\texttt{upon}}}\;\mathrm{{\mathtt{upon}}}\;}
\def\masa{\index{asa@{\texttt{asa}}}\;\mathrm{{\mathtt{asa}}}\;}

\def\Tfirst{\index{first@{\texttt{first}}}\texttt{first}}
\def\Tnext{\index{next@{\texttt{next}}}\texttt{next}}
\def\Tfby{\index{fby@{\texttt{fby}}}\texttt{fby}}
\def\Twvr{\index{wvr@{\texttt{wvr}}}\texttt{wvr}}
\def\Tupon{\index{upon@{\texttt{upon}}}\texttt{upon}}
\def\Tasa{\index{asa@{\texttt{asa}}}\texttt{asa}}

\newcommand{\eqdef}{\stackrel{{\mathrm{def}}}{=}}

\newcommand{\myabstract}
{
A Forensic Lucid intensional programming language has been proposed for
intensional cyberforensic analysis.
In large part, the language is based on
various
predecessor and codecessor Lucid dialects 
bound by the
higher-order intensional logic (HOIL) that is behind them. This work formally specifies
the operational
aspects
of the Forensic Lucid language and compiles a
theory of its
constructs
using Isabelle, a proof assistant system.
}

\lstdefinestyle{codeStyle}
{
	language=Java,
	frame=single,  
	basicstyle=\footnotesize,
	captionpos=b,
	showstringspaces=false,
	showspaces=false,
	extendedchars=true,
	linewidth=1\linewidth,
	breaklines=true,
	float=phtb  
}

\renewcommand{\textfloatsep}{.005pt} 

\begin{document}

\title{Formally Specifying and Proving Operational Aspects\\of {\sf Forensic Lucid} in {\sf Isabelle}}
\titlerunning{Formally Specifying and Proving Operational Aspects of {\sf Forensic Lucid} in {\sf Isabelle}}

\author{Serguei A. Mokhov and Joey Paquet}
\institute{Department of Computer Science and Software Engineering\\Faculty of Engineering and Computer Science\\Concordia University, Montr\'eal, Qu\'ebec, Canada,\\\email{\{mokhov,paquet\}@cse.concordia.ca}
}

\maketitle

\begin{abstract}
{\myabstract}
\end{abstract}

\section{Introduction}

As a part of the Intensional Cyberforensics project, we define a functional-intensional
programming/specification language, called {\flucid}. The language is under active design and development
including its syntax, semantics, the corresponding compiler, run-time, and
interactive ``development''
environments~\cite{mokhov-iforensics-tech07,mokhov-iforensics-phdproposal07} that
we refer to as General Intensional Programming System ({\gipsy})~\cite{gipsy}.
We approach the problem using Isabelle~\cite{isabelle} as a proof assistant.

\paragraph{Problem Statement.}

A lot of intensional dialects have been spawned from the
functional intensional
programming language called {\lucid}~\cite{lucid85,lucid95,lucid76,lucid77,nonprocedural-iterative-lucid-77,rlucid99,paquetThesis,crr05}.
{\lucid} (see~\xs{sect:lucid-summary}) itself was invented with
a goal for program correctness verification~\cite{lucid76,lucid77}.
While there were a number of
operational semantics rules for compiler and run-time environments developed for
all those dialects throughout the years, there was no a complete formal proof set
of the rules of the languages. Yet another dialect of {\lucid} has been created
to foster the research on intensional cyberforensics (see~\xs{sect:cyberforensics-summary}), called {\flucid}, which,
in a large part is a union of the syntax and operational semantics rules from
the comprising languages with the forensic extensions. In order to be a credible
tool to use, for example, in court, to implement relevant tools for the argumentation, the language ought to have a solid
scientific base, a part of which is formalizing the semantics the language and
proving correctness of the programs written in it.

\paragraph{Proposed Solution.}

In this work, we propose to begin validation of the
{\flucid} constructs with the
Isabelle prover assistant~\cite{isabelle} and extend it to the comprising
Lucid dialects as a whole. We proceed bottom-up from ``core'' Lucid dialects
such as {\gipl}, {\lucx}, and {\ilucid} and even their smaller decompositions
as well as top-down from {\flucid}
to arrive to a comprehensive set of proofs covering the dialects.

\subsection{Intensional Logics and Programming}
\index{intensional!programming}
\index{intensional!logic}

\subsubsection{Definitions.}

Intensional programming (IP) is based on intensional (or multidimensional)
logics\index{intensional!logic}\index{logic!intensional}, which, in turn, are based on natural language understanding
aspects (such as time, belief, situation, and direction).
IP brings in {\bf dimensions}\index{dimensions} and {\bf context}\index{context} to programs (e.g. space and time
in physics or chemistry). Intensional logic adds dimensions to logical
expressions; thus, a non-intensional logic\index{logic!non-intensional} can be seen as a constant or a
snapshot in all possible dimensions. {\it Intensions are dimensions} at which a
certain statement is true or false (or has some other than a Boolean value).
{\it Intensional operators}\index{intensional!operators} are operators that allow us to navigate within these
dimensions. {\em Higher-order intensional logic} (HOIL) is the one that couples functional
programming as that of {\lucid} with multidimensional dataflows that the intensional
programs can query an alter through an explicitly notion of contexts as first-class
values~\cite{wanphd06,gipsy-simple-context-calculus-08}.

\subsubsection{An Example of Using Temporal Intensional Logic.}
\index{logic!temporal}

Temporal intensional logic is an extension of temporal logic that allows
to specify the time in the future or in the past.

(1) \tab{20} $E_1$ := it is raining {\bf here} {\bf today}

Context: \{\texttt{place:}{\bf here}, \texttt{time:}{\bf today}\}

(2) \tab{20} $E_2$ := it was raining {\bf here} {\it before}({\bf today}) = {\it yesterday}

(3) \tab{20} $E_3$ := it is going to rain {\it at} (altitude {\bf here} + 500 m) {\it after}({\bf today}) = {\it tomorrow}

Let's take $E_1$ from (1) above. Then
let us fix {\bf here} to {\bf Montreal} and assume it is a {\it constant}.
In the month of February, 2008, with granularity of day, for every day, we can
evaluate $E_1$ to either {\it true} or {\it false}:

\begin{verbatim}
Tags:   1 2 3 4 5 6 7 8 9 ...
Values: F F T T T F F F T ...
\end{verbatim}

If one starts varying the {\bf here} dimension (which could even be broken down
to $X$, $Y$, $Z$), one gets a two-dimensional evaluation of $E_1$:

\begin{verbatim}
City: /   1 2 3 4 5 6 7 8 9 ...
Montreal  F F T T T F F F T ...
Quebec    F F F F T T T F F ...
Ottawa    F T T T T T F F F ...
\end{verbatim}

\subsection{Lucid}
\label{sect:lucid-summary}

{\lucid}~\cite{lucid85,lucid95,nonprocedural-iterative-lucid-77,lucid76,lucid77} is a dataflow intensional and functional programming language. In fact, it is a family of languages that are built upon intensional logic (which in turn can be understood as a multidimensional generalization of temporal logic) involving context and demand-driven parallel computation model. A program written in some {\lucid} dialect is an expression that may have subexpressions that need to be evaluated at certain {\it context}. Given the set of dimension $D=\{dim_i\}$ in which an expression varies, and a corresponding set of indexes or {\it tags} defined as placeholders over each dimension, the context is represented as a set of $<\!\!dim_i:tag_i\!\!>$ mappings and each variable in {\lucid}, called often a {\em stream}, is evaluated in that defined context that may also evolve using context operators~\cite{gipsy-simple-context-calculus-08,tongxinmcthesis08,kaiyulucx,wanphd06}. The generic version of {\lucid}, {\gipl}~\cite{paquetThesis}, defines two basic operators \api{@} and \api{\#} to navigate in the contexts (switch and query).
The {\gipl} was the first
generic programming language of all intensional languages, defined by the means
of only two intensional operators \api{@} and \api{\#}. It has been proven that other intensional programming languages of the Lucid family can be translated into the {\gipl}~\cite{paquetThesis}.
Please refer to \xa{appdx:lucid-proofs} for the greater details about Lucid
origins, variables as streams, random access to streams, and the basic operators.
Since the {\lucid} family of language thrived around intensional logic that
makes the notion of context explicit and central, and recently, a first class
value~\cite{kaiyulucx,wanphd06,gipsy-simple-context-calculus-08,tongxinmcthesis08}
that can be passed around as function parameters or as return values
and have a set of operators defined upon. We greatly draw on this
notion by formalizing our evidence and the stories as a contextual
specification of the incident to be tested for consistency against
the incident model specification. In our specification model we require
more than just atomic context values -- we need a higher-order
context hierarchy to specify different level of detail of the
incident and being able to navigate into the ``depth'' of such
a context. A similar provision by has already been made by the
author~\cite{marfl-context-secasa08} and earlier works of Swoboda et al.
in~\cite{swobodaphd04,intensionalisation-tools,distributed-context-computing,active-functional-idatabase}
that needs some modifications to the expressions of the cyberforensic
context.

Some other languages can be referred to as intensional even though they may not
refer to themselves as such, and were born after {\lucid} ({\lucid} began in 1974).
Examples include hardware-description languages (HDLs, appeared in 1977) where the notion of
time (often the only ``dimension'', and usually progresses only forward), e.g.
Verilog and VHDL. Another branch of newer languages for the becoming popular
is aspect-oriented programming (AOP) languages, that can have a notion of
context explicitly, but primarily focused on software engineering aspect of
software evolution and maintainability.

\subsection{Cyberforensic Analysis}
\label{sect:cyberforensics-summary}

Cyberforensic analysis has to do with automated or semi-automated
processing of and reasoning about electronic evidence, witnesses, and other details
from cybercrime incidents (involving computers, but not limited
to them). Analysis is one of the phases in cybercrime investigation,
where the others focus on evidence collection, preservation,
chain of custody, information extraction that precede the analysis.
The phases the follow the analysis are formulation of a report
and potential prosecution, typically involving expert witnesses.
There are quite a few techniques, tools (hardware and software),
and methodologies have been developed for all the briefly mentioned
phases of the cybercrime investigation. A lot of attention has
been paid to the tool development for evidence collection
and preservation; a few tools have been developed to aid ``browsing''
data in the confiscated storage media, log files, memory, and so
on. A lot less number of tools have been developed for case analysis
of the data, and the existing commercial packages (e.g. Encase or
FTK) are very expensive. Even less so there are case management,
event modeling, and event reconstruction, especially with solid
formal theoretical base. The first formal approach to the cybercrime
investigation was the finite-state automata (FSA) approach by
Gladyshev et. al~\cite{printer-case,blackmail-case}. The approach
is complex to use and understand for non computer science or equivalent
investigators. The aim of {\flucid} is to alleviate those difficulties,
be sound and complete, expressive and usable, and provide even
further usability improvement with the graphic interface that allow
data-flow graph-based (DFG) programming that allows translation between
DFGs and Lucid code for compilation and is implemented for {\ilucid}
in {\gipsy} already~\cite{yimin04}, and requires forensic extensions.
While {\flucid} is in the design and implementation, its solid base
is being established in part with this work. The goal of {\flucid}
in the cyberforensic analysis is to be able to express in a program
form the encoding of the evidence, witness stories, and evidential
statements, that can be tested against claims to see if there is
a possible sequence or multiple sequences of events that explain
a given story. This is designed to aid investigator to avoid ad-hoc
conclusions and have them look at the possible explanations the {\flucid}
program execution would yield and refine the investigation, as
was shown in the works~\cite{printer-case,blackmail-case} investigators
failed to analyze all the stories and their plausibility before
drawing conclusions in the case. We do not recite the cases here
due to the length limitations.

%
%

\section{{\flucid}}
\label{sect:flucid}
\label{sect:forensic-lucid}

The end goal is to define our {\flucid} language where its constructs concisely
express cyberforensic evidence, which can be initial state of
a
case
towards what we have actually observed as a final state.
The implementing system (i.e. {\gipsy}) has to backtrace intermediate results in order to provide
the corresponding event reconstruction path, if it exists.
The result of the expression in its basic form is either {\it true} or {\it false},
i.e. ``guilty'' or ``not guilty'' given the context per explanation with the backtrace.
There can be multiple backtraces, that correspond to the explanation of the evidence (or lack thereof).

\subsection{Properties}
\index{{\flucid}!Features}
\index{{\flucid}!Properties}

We define {\flucid} to model the evidential statements and other expressions representing
the evidence and observations as a higher-order context hierarchy.
An execution trace of a {\flucid} program would expose the possibility of the proposed
claim with the events in the middle.

Addition of the context calculus from {\lucx} for operators on {\lucx}'s context sets
(\lucidop{union}, \lucidop{intersection}, etc.) are used to address to provide
a collection of traces. {\flucid} inherits the properties of
{\lucx}, {\marfl}, {\olucid}, {\jooip}
(and their comprising dialects),
where the former is for the context calculus, and the latter for the arrays
and structural representation of data for modeling the case data structures
such as events, observations, and groupings of the related data.

One of the basic requirements is that the complete definition of
the operational semantics of {\flucid} should be compatible with the
basic {\lucx} and {\gipl}, i.e. the translation rules or equivalent are
to be provided when implementing the language compiler within {\gipsy},
and such that the {\gee} can execute it with minimal changes.

\begin{lstlisting}[
    label={list:story-board-expression},
    caption={Intensional Storyboard Expression},
    style=codeStyle
    ]
foo @
{
  [ final observed event, possible initial observed event ],
  [            ],
  [            ]
}
\end{lstlisting}

While the \verb+[...]+ notation here may be confusing with respect to the notation of \texttt{[dimension:tag]}
in {\lucid} and more specifically in {\lucx}~\cite{wanphd06,gipsy-context-calculus-07}, it is in fact a simple
syntactical extension to allow higher-level groups of contexts where this syntactical sugar is later translated
to the baseline context constructs.
The tentative notation of \verb+{[...],...,[...]}+ implies a notion similar to the notion of the ``context set''
in~\cite{wanphd06,gipsy-context-calculus-07} except with the syntactical sugar mentioned earlier where we allow
syntactical grouping of properties, observations, observation sequences, and evidential statements as our context sets.

\subsection{Transition Function}
\label{sect:tans-func}
\index{Transition Function}
\index{{\flucid}!Transition Function}

A transition function determines how the context of evaluation changes
during computation. A general issue exists that we have to address
is that the transition function {\trans} is problem-specific.
In the FSA approach, the transition function is the labeled graph itself. In the first prototype,
we follow the graph to model our {\flucid} equivalent.
In general,
Lucid has already basic operators to navigate and switch from one context to another,
which represent the basic transition functions in themselves (the intensional
operators such as \api{@}, \api{\#}, \api{iseod}, \api{first}, \api{next}, \api{fby}, \api{wvr}, \api{upon},
and \api{asa} as well as their inverse operators\footnote{Defined further.}).
However, a specific problem being modeled requires more specific transition function
than just plain intensional operators. In this case the transition function is
a {\flucid} function where the matching state transition modeled through
a sequence of intensional operators.
In fact, the forensic operators are just pre-defined functions that rely on traditional
and inverse Lucid operators as well as context switching operators
that achieve something similar to the transitions in~\cite{printer-case,blackmail-case}.
In fact, the intensional operators
of {\lucid} represent the basic building blocks for {\trans} and {\invtrans}.

\subsection{Primitive Operators}
\index{Forensic Lucid!New Operators}
\index{Forensic Lucid!Operators}
\index{Operators!Forensic Lucid}

The basic set of the classic intensional operators is extended with the similar
operators, but inverted in one of their aspects: either negation of trueness or reverse of direction of navigation.
Here we provide an informal definition followed by their formal counterpart
of these operators alongside with the classical ones (to
remind the reader what they do and enlighten the unaware reader). The reverse operators have a restriction that they must work on
the bounded streams at the positive infinity. This is not a stringent limitation as the
our contexts of observations and evidence in this work are always finite, so they all have
the beginning and the end. What we need is an ability to go back in the stream and, perhaps,
negate in it with classical-like operators, but reversed.

The
operators
are
defined below to give
a complete picture. The classical operators
\olucidop{first}, \olucidop{next}, \olucidop{fby},
\olucidop{wvr}, \olucidop{upon}, and \olucidop{asa}
were previously defined in~\cite{paquetThesis} and
earlier. The other complimentary, inverse, and
negation operators were defined and revised from~\cite{mokhov-cyberforensics-07}.
In this list of operators, especially the reverse ones, we make an
important assumption that the streams we are working with are finite,
which is sufficient for our tasks.
Thus, our streams of context values can be bound between \api{bod}
and \api{eod} and contain a finite tag set of elements is used as
a context type.
For summary of the application of the just defined operators' examples,
please refer to \xa{appdx:new-operators-table}.

\index{Forensic Lucid!Operators @ and \#}

Following the steps in~\cite{paquetThesis}, we further represent the definition of the
operators via \lucidop{@} and \lucidop{\#}.
Again, there is a mix of classical operators that were previously
defined in~\cite{paquetThesis}, such as \lucidop{first}, \lucidop{next},
\lucidop{fby}, \lucidop{wvr}, \lucidop{upon}, and \lucidop{asa} as well
as the new operators from this work. The collection of the translated operators
denoted in
\lucidop{monospaced font}, while we provide their equivalence to the original Lucid
operators, denoted as \olucidop{small caps}.

The primitive operators are founding blocks to construct more complex
case-specific functions that represent a particular investigation case
as well as more complex so-called {\em forensic operators}.

\begin{itemize}

\item A stream of first elements of stream $X$:

	\olucidop{first} $X = (x_0, x_0, ..., x_0, ...)$

\begin{equation}
\lucidop{first}\;\;X = X @ 0
\label{eq:first}
\end{equation}

\item A stream of second elements of stream $X$:

	\olucidop{second} $X = (x_1, x_1, ..., x_1, ...) = $ \olucidop{first} \olucidop{next} $X$

\item A stream of last elements of stream $X$:

	\olucidop{last} $X = (x_n, x_n, ..., x_n, ...)$

\noindent
This definition of the \olucidop{last} operator relies on the earlier
stated assumption that our streams can be explicitly finite for the language
we are developing. This affects the follow up operators that rely in that fact
just as well. It is also important to note that the \olucidop{last} operator
in our design does not return \api{eod} all the time on the finite stream
due to lack of usefulness for such a value; instead it returns the element
of the stream just before the \api{eod}.

\begin{equation}
\lucidop{last}\;\;X = X @ (\# @ (\#\lucidop{iseod}(\#) - 1))
\end{equation}

\item A stream of elements one before the last one of stream $X$:

	\olucidop{prelast} $X = (x_{n-1}, x_{n-1}, ..., x_{n-1}, ...) = $ \olucidop{last} \olucidop{prev} $X$

\item A stream of elements of stream $X$ other than the first:

	\olucidop{next} $X = (x_1, x_2, ..., x_{i+1}, ...)$

\begin{equation}
\lucidop{next}\;\;X = X @ (\# + 1)
\end{equation}

\item A stream of elements of stream $X$ other than the last:

	\olucidop{prev} $X = (x_{n-1}, ..., x_{i+1}, x_i, x_{i-1}, ...)$
	
\begin{equation}
\lucidop{prev}\;\;X = X @ (\# - 1)
\end{equation}

\item First element of $X$ followed by all of $Y$:

	$X$ \olucidop{fby} $Y = (x_0, y_0, y_1, ..., y_{i-1}, ...)$

\begin{eqnarray}
X\;\;\lucidop{fby}\;\;Y &=& \mathbf{if}\;\# = 0\;\mathbf{then}\;X\;\mathbf{else}\;Y @ (\# - 1)\\\nonumber
                        &=& \mathbf{if}\;\lucidop{isbod}\;X\;\mathbf{then}\;X\;\mathbf{else}\;\lucidop{prev}\;Y
\end{eqnarray}

\item First element of $X$ preceded by all of $Y$:

	$X$ \olucidop{pby} $Y = (y_0, y_1, ..., y_{i-1}, ..., y_n, x_0)$

\begin{eqnarray}
X\;\;\lucidop{pby}\;\;Y &=& \mathbf{if}\;\lucidop{iseod}\;\#\;\mathbf{then}\;X\;\mathbf{else}\;Y @ (\# + 1)\\\nonumber
                        &=& \mathbf{if}\;\lucidop{iseod}\;Y\;\mathbf{then}\;X\;\mathbf{else}\;\lucidop{next}\;Y
\end{eqnarray}

\item Stream of negated arithmetic values of $X$:

	\olucidop{neg} $X = (-x_0, -x_1, -x_2, ..., -x_{i+1}, ...)$

\begin{equation}
\lucidop{neg}\;\;X = -X\\
\end{equation}

\item Stream of inverted truth values of $X$:

	\olucidop{not} $X = (!x_0, !x_1, !x_2, ..., !x_{i+1}, ...)$

\begin{equation}
\lucidop{not}\;\;X = \mathbf{if}\;X\;\mathbf{then}\;!X\;\mathbf{else}\;X
\end{equation}

\item A logical AND stream of truth values of $X$ and $Y$:

	$X$ \olucidop{and} $Y = (x_0 \&\& y_0, x_1 \&\& y_1, x_2 \&\& y_2, ..., x_{i+1} \&\& y_{i+1}, ...)$

\begin{equation}
X\;\;\lucidop{and}\;\;Y = X \&\& Y
\end{equation}

\item A logical OR stream of truth values of $X$ and $Y$:

	$X$ \olucidop{or} $Y = (x_0 || y_0, x_1 || y_1, x_2 || y_2, ..., x_{i+1} || y_{i+1}, ...)$

\begin{equation}
X\;\;\lucidop{or}\;\;Y = X || Y\\
\end{equation}

\item A logical XOR stream of truth values of $X$ and $Y$:

	$X$ \olucidop{xor} $Y = (x_0 \oplus y_0, x_1 \oplus y_1, x_2 \oplus y_2, ..., x_{i+1} \oplus y_{i+1}, ...)$

\begin{equation}
X\;\;\lucidop{xor}\;\;Y = \lucidop{not} ((X\;\lucidop{and}\;Y)\;\lucidop{or}\;\lucidop{not}\;(X\;\lucidop{or}\;Y))
\end{equation}

\item \olucidop{wvr} stands for {\it whenever}. \olucidop{wvr} chooses
from its left-hand-side operand only values in the current dimension where
the right-hand-side evaluates to {\em true}.

	$X$ \olucidop{wvr} $Y =$

	\tab{20} {\bf if} \olucidop{first} $Y \neq 0$

	\tab{20} {\bf then} $X$ \olucidop{fby} (\olucidop{next} $X$ \olucidop{wvr} \olucidop{next} $Y)$

	\tab{20} {\bf else} (\olucidop{next} $X$ \olucidop{wvr} \olucidop{next} $Y)$

\begin{eqnarray}
X\;\;\lucidop{wvr}\;\;Y &=& X @ T\;\lucidop{where}\\\nonumber
                        & & \tab{20} T = U\;\lucidop{fby}\;U @ (T + 1)\\\nonumber
                        & & \tab{20} U = \mathbf{if}\;Y\;\mathbf{then}\;\#\;\mathbf{else}\;\lucidop{next}\;U\\\nonumber
                        & & \lucidop{end}
\end{eqnarray}

\item \olucidop{rwvr} stands for {\it retreat whenever}. \olucidop{rwvr} chooses
from its left-hand-side operand backwards only values in the current dimension where
the right-hand-side evaluates to {\em true}.

	$X$ \olucidop{rwvr} $Y =$

	\tab{20} {\bf if} \olucidop{last} $Y \neq 0$

	\tab{20} {\bf then} $X$ \olucidop{pby} (\olucidop{prev} $X$ \olucidop{rwvr} \olucidop{prev} $Y)$

	\tab{20} {\bf else} (\olucidop{prev} $X$ \olucidop{rwvr} \olucidop{prev} $Y)$

\begin{eqnarray}
X\;\;\lucidop{rwvr}\;\;Y &=& X @ T\;\lucidop{where}\\\nonumber
                        & & \tab{20} T = U\;\lucidop{pby}\;U @ (T - 1)\\\nonumber
                        & & \tab{20} U = \mathbf{if}\;Y\;\mathbf{then}\;\#\;\mathbf{else}\;\lucidop{prev}\;U\\\nonumber
                        & & \lucidop{end}
\end{eqnarray}

\item \olucidop{nwvr} stands for {\it not whenever}. \olucidop{nwvr} chooses
from its left-hand-side operand only values in the current dimension where
the right-hand-side evaluates to {\em false}.

	$X$ \olucidop{nwvr} $Y = X$ \olucidop{wvr} \olucidop{not} $Y =$

	\tab{20} {\bf if} \olucidop{first} $Y == 0$

	\tab{20} {\bf then} $X$ \olucidop{fby} (\olucidop{next} $X$ \olucidop{nwvr} \olucidop{next} $Y)$

	\tab{20} {\bf else} (\olucidop{next} $X$ \olucidop{nwvr} \olucidop{next} $Y)$

\begin{eqnarray}
X\;\;\lucidop{nwvr}\;\;Y &=& X @ T\;\lucidop{where}\\\nonumber
                        & & \tab{20} T = U\;\lucidop{fby}\;U @ (T + 1)\\\nonumber
                        & & \tab{20} U = \mathbf{if}\;Y == 0\;\mathbf{then}\;\#\;\mathbf{else}\;\lucidop{next}\;U\\\nonumber
                        & & \lucidop{end}
\end{eqnarray}

\item \olucidop{nrwvr} stands for {\it do not retreat whenever}. \olucidop{nrwvr} chooses
from its left-hand-side operand backwards only values in the current dimension where
the right-hand-side evaluates to {\em false}.

	$X$ \olucidop{nrwvr} $Y = X$ \olucidop{rwvr} \olucidop{not} $Y =$

	\tab{20} {\bf if} \olucidop{last} $Y == 0$

	\tab{20} {\bf then} $X$ \olucidop{pby} (\olucidop{prev} $X$ \olucidop{nrwvr} \olucidop{prev} $Y)$

	\tab{20} {\bf else} (\olucidop{prev} $X$ \olucidop{nrwvr} \olucidop{prev} $Y)$

\begin{eqnarray}
X\;\;\lucidop{rnwvr}\;\;Y &=& X @ T\;\lucidop{where}\\\nonumber
                        & & \tab{20} T = U\;\lucidop{pby}\;U @ (T - 1)\\\nonumber
                        & & \tab{20} U = \mathbf{if}\;Y == 0\;\mathbf{then}\;\#\;\mathbf{else}\;\lucidop{prev}\;U\\\nonumber
                        & & \lucidop{end}
\end{eqnarray}

\item \olucidop{asa} stands for {\it as soon as}. \olucidop{asa} returns
the value of its left-hand-side as a first point in that stream as soon as the
right-hand-side evaluates to {\em true}.

	$X$ \olucidop{asa} $Y =$ \olucidop{first} $(X$ \olucidop{wvr} $Y)$

\begin{equation}
X\;\;\lucidop{asa}\;\;Y = \lucidop{first}\;(X\;\lucidop{wvr}\;Y)
\end{equation}

\item \olucidop{ala} (other suggested name is \olucidop{rasa}) stands for {\it as late as} (or {\it reverse of a soon as}). \olucidop{ala} returns
the value of its left-hand-side as the last point in that stream when the
right-hand-side evaluates to {\em true} for the last time.

	$X$ \olucidop{ala} $Y =$ \olucidop{last} $(X$ \olucidop{wvr} $Y)$

\begin{equation}
X\;\;\lucidop{ala}\;\;Y = \lucidop{last}\;(X\;\lucidop{rwvr}\;Y)\\
\end{equation}

\item \olucidop{nasa} stands for {\it not as soon as}. \olucidop{nasa} returns
the value of its left-hand-side as a first point in that stream as soon as the
right-hand-side evaluates to {\em false}.

	$X$ \olucidop{nasa} $Y =$ \olucidop{first} $(X$ \olucidop{nwvr} $Y)$

\begin{equation}
X\;\;\lucidop{nasa}\;\;Y = \lucidop{first}\;(X\;\lucidop{nwvr}\;Y)\\
\end{equation}

\item \olucidop{nala} (other suggested name is \olucidop{nrasa}) stands for {\it not as late as} (or {\it reverse of not a soon as}).
\olucidop{nala} returns the value of its left-hand-side as the last point in that stream when the
right-hand-side evaluates to {\em false} for the last time.

	$X$ \olucidop{nala} $Y =$ \olucidop{last} $(X$ \olucidop{nwvr} $Y)$

\begin{equation}
X\;\;\lucidop{nala}\;\;Y = \lucidop{last}\;(X\;\lucidop{nrwvr}\;Y)\\
\end{equation}

\item \olucidop{upon} stands for {\it advances upon}. Unlike \olucidop{asa}, \olucidop{upon}
switches context of its left-hand-side operand if the right-hand side
is {\em true}.

	$X$ \olucidop{upon} $Y = X$ \olucidop{fby} $($

	\tab{20} {\bf if} \olucidop{first} $Y \neq 0$

	\tab{20} {\bf then} $($\olucidop{next} $X$ \olucidop{upon} \olucidop{next} $Y)$

	\tab{20} {\bf else} $(X$ \olucidop{upon} \olucidop{next} $Y))$

\begin{eqnarray}
X\;\;\lucidop{upon}\;\;Y &=& X @ W\;\lucidop{where}\\\nonumber
                        & & \tab{20} W = 0\;\lucidop{fby}\;(\mathbf{if}\;Y\;\mathbf{then}\;(W + 1)\;\mathbf{else}\;W)\\\nonumber
                        & & \lucidop{end}
\end{eqnarray}

\item \olucidop{rupon} stands for {\it retreats upon}. \olucidop{rupon}
switches context backwards of its left-hand-side operand if the right-hand side
is {\em true}.

	$X$ \olucidop{rupon} $Y = X$ \olucidop{pby} (

	\tab{20} {\bf if} \olucidop{last} $Y \neq 0$

	\tab{20} {\bf then} $($\olucidop{prev} $X$ \olucidop{rupon} \olucidop{prev} $Y)$

	\tab{20} {\bf else} $(X$ \olucidop{rupon} \olucidop{prev} $Y))$

\begin{eqnarray}
X\;\;\lucidop{rupon}\;\;Y &=& X @ W\;\lucidop{where}\\\nonumber
                        & & \tab{20} W = 0\;\lucidop{pby}\;(\mathbf{if}\;Y\;\mathbf{then}\;(W - 1)\;\mathbf{else}\;W)\\\nonumber
                        & & \lucidop{end}
\end{eqnarray}

\item \olucidop{nupon} stands for {\it not advances upon} or rather {\it advances otherwise}. \olucidop{nupon}
switches context of its left-hand-side operand if the right-hand side
is {\em false}.

	$X$ \olucidop{nupon} $Y = X$ \olucidop{upon} \olucidop{not} $Y = X$ \olucidop{fby} (

	\tab{20} {\bf if} \olucidop{first} $Y == 0$

	\tab{20} {\bf then} $($\olucidop{next} $X$ \olucidop{nupon} \olucidop{next} $Y)$

	\tab{20} {\bf else} $(X$ \olucidop{nupon} \olucidop{next} $Y))$

\begin{eqnarray}
X\;\;\lucidop{nupon}\;\;Y &=& X @ W\;\lucidop{where}\\\nonumber
                        & & \tab{20} W = 0\;\lucidop{fby}\;(\mathbf{if}\;Y == 0\;\mathbf{then}\;(W + 1)\;\mathbf{else}\;W)\\\nonumber
                        & & \lucidop{end}
\end{eqnarray}

\item \olucidop{nrupon} stands for {\it not retreats upon}. \olucidop{nrupon}
switches context backwards of its left-hand-side operand if the right-hand side
is {\em false}.

	$X$ \olucidop{nrupon} $Y = X$ \olucidop{rupon} \olucidop{not} $Y = X$ \olucidop{pby} (

	\tab{20} {\bf if} \olucidop{last} $Y == 0$

	\tab{20} {\bf then} $($\olucidop{prev} $X$ \olucidop{nrupon} \olucidop{prev} $Y)$

	\tab{20} {\bf else} $(X$ \olucidop{nrupon} \olucidop{prev} $Y))$

\begin{eqnarray}
X\;\;\lucidop{nrupon}\;\;Y &=& X @ W\;\lucidop{where}\\\nonumber
                        & & \tab{20} W = 0\;\lucidop{pby}\;(\mathbf{if}\;Y == 0\;\mathbf{then}\;(W - 1)\;\mathbf{else}\;W)\\\nonumber
                        & & \lucidop{end}
\end{eqnarray}

\end{itemize}

\subsection{Forensic Operators}

The operators presented here are based on the discussion of the combination
function and others that form more-than-primitive
operations to support the required implementation.
The discussed earlier \api{comb()} operator needs to be realized in the general manner for combining analogies of MPRs,
which in our case are higher-level contexts, in the new language's dimension types.

\begin{itemize}

\item 
\lucidop{combine} corresponds to the $comb$ function as originally described by Gladyshev in~\cite{printer-case}.
It is defined in \xl{list:combine}. It is a preliminary context-enhanced version.

\begin{lstlisting}[
    label={list:combine},
    caption={The \lucidop{combine} Operator},
    style=codeStyle
    ]
/**
 * Append given e to each element
 * of a given stream e under the
 * context of d.
 *
 * @return the resulting combined stream
 */
combine(s, e, d) =
  if iseod s then eod;
  else (first s fby.d e) fby.d combine(next s, e, d);
  fi
\end{lstlisting}

\item
\lucidop{product} tentatively corresponds to the cross-product~\cite{printer-case} of contexts.
It is defined in \xl{list:product}.

\begin{lstlisting}[
    label={list:product},
    caption={The \lucidop{product} Operator},
    style=codeStyle
    ]
/**
 * Append elements of s2 to element of s1
 * in all possible combinations.
 */
product(s1, s2, d) =
  if iseod s2 then eod;
  else combine(s1, first s2) fby.d product(s1, next s2);
  fi
\end{lstlisting}

\end{itemize}

The translated examples show recursion that we are not prepared to deal with
in the current Lucid semantics, and will address that in the future work.
The two illustrated operators are the first of the a few more to follow
in the final language prototype.


\subsection{Operational Semantics}
\label{appdx:semantics}
\label{sect:semantics}
\index{Operational Semantics}
\index{Forensic Lucid!Operational Semantics}
\index{Operational Semantics!Forensic Lucid}
\index{Operational Semantics!Indexical Lucid}
\index{Operational Semantics!Lucx}

As previously mentioned, the operational semantics of {\flucid} for the large
part is viewed as a composition of the semantic rules of {\ilucid}, {\olucid}, and {\lucx}
along with the new operators and definitions. Here we list the existing combined
semantic definitions to be used the new language, specifically extracts of
operational semantics from {\gipl}~\cite{paquetThesis},
and {\lucx}~\cite{wanphd06} are in \xf{fig:ilucid-semantics},
and \xf{fig:lucx-semantics} respectively. The explanation of the rules and the notation
are given in great detail in the cited works and are trimmed in this article. For convenience
of the reader they are recited here to a degree.
The new rules of the operational semantics of {\flucid} cover the newly defined
operators primarily, including the reverse and logical stream operators as well
as forensic-specific operators.
We use the same
notation as the referenced languages to maintain consistency in defining our
rules.

\begin{figure}[htb!]
\begin{footnotesize}
\begin{eqnarray}
{\mathbf{E_{cid}}} &:& \frac
  {\johndef(\myid)=(\texttt{const},c)} {\context{\myid}{c}}\label{sem:gipl:e.cid}\\\nonumber\\
{\mathbf{E_{opid}}} &:& \frac
  {\johndef(\myid)=(\texttt{op},f)}
  {\context{\myid}{\myid}}\label{sem:gipl:e.opid}\\\nonumber\\
{\mathbf{E_{did}}} &:& \frac
  {\johndef(\myid)=(\texttt{dim})}
  {\context{\myid}{\myid}}\label{sem:gipl:e.did}\\\nonumber\\
{\mathbf{E_{fid}}} &:& \frac
  {\johndef(\myid)=(\texttt{func},\myid_i,E)}
  {\context{\myid}{\myid}}\label{sem:gipl:e.fid}\\\nonumber\\
{\mathbf{E_{vid}}} &:& \frac
  {\johndef(\myid)=(\texttt{var},E)\qquad
   \context{E}{v}}
  {\context{\myid}{v}}\label{sem:gipl:e.vid}\\\nonumber\\
{\mathbf{E_{op}}} &:& \frac
   {\context{E}{\myid}\qquad
    \johndef(\myid)=(\texttt{op},f)\qquad
    \context{E_i}{v_i}
   }
   {\context{E(E_1,\ldots,E_n)}{f(v_1,\ldots,v_n)}}\label{sem:gipl:e.op}\\\nonumber\\
{\mathbf{E_{fct}}} &:& \frac
   {\context{E}{\myid}\qquad
    \johndef(\myid)=(\texttt{func},\myid_i,E')\qquad
    \context{E'[\myid_i\leftarrow E_i]}{v}
   }
   {\context{E(E_1,\ldots,E_n)}{v}}\label{sem:gipl:e.fct}\\\nonumber\\
{\mathbf{E_{c_T}}} &:& \frac
  {\context{E}{\textit{true}}\qquad
   \context{E'}{v'}
  }
  {\context{\myifthenelse}{v'}}\label{sem:gipl:e.ct}\\\nonumber\\
{\mathbf{E_{c_F}}} &:& \frac
  {\context{E}{\textit{false}}\qquad
   \context{E''}{v''}
  }
  {\context{\myifthenelse}{v''}}\label{sem:gipl:e.cf}\\\nonumber\\
{\mathbf{E_{tag}}} &:& \frac
  {\context{E}{\myid}\qquad
   \johndef(\myid)=(\texttt{dim})
  }
  {\context{\#E}{\mathcal{P}(\myid)}}\label{sem:gipl:e.tag}\\\nonumber\\
{\mathbf{E_{at}}} &:& \frac
  {\context{E'}{\myid}\qquad
   \johndef(\myid)=(\texttt{dim})\qquad
   \context{E''}{v''}\qquad
   \mathcal{D},\mathcal{P}\mydagger[\myid\mapsto v''] \vdash E : v
  }
  {\context{E\;@E'\;E''}{v}}\label{sem:gipl:e.at}\\\nonumber\\
{\mathbf{E_{w}}} &:& \frac
  {\qcontext{Q}{\mathcal{D}',\mathcal{P}'}\qquad
   \mathcal{D}',\mathcal{P}' \vdash E : v
  }
  {\context{E\;\mathtt{where}\;Q}{v}}\label{sem:gipl:e.w}\\\nonumber\\
{\mathbf{Q_{dim}}} &:& \frac
  {}
  {\qcontext{\texttt{dimension}\;\myid}
   {\mathcal{D}\mydagger[\myid\mapsto(\texttt{dim})],
   \mathcal{P}\mydagger[\myid\mapsto 0]}
  }\label{sem:gipl:q.dim}\\\nonumber\\
{\mathbf{Q_{id}}} &:& \frac
  {}
  {\qcontext{\myid=E}
   {\mathcal{D}\mydagger[\myid\mapsto(\texttt{var},E)],
    \mathcal{P}}
  }\label{sem:gipl:q.id}\\\nonumber\\
{\mathbf{Q_{fid}}} &:& \frac
   {}
   {\qcontext{\myid(\myid_1,\ldots,\myid_n)=E}
    {\mathcal{D}\mydagger[\myid\mapsto(\texttt{func},\myid_i,E)],
    \mathcal{P}}
   }\label{sem:gipl:q.fid}\\\nonumber\\
{\mathbf{QQ}} &:& \frac
  {\qcontext{Q}{\mathcal{D}',\mathcal{P}'}\qquad
   \mathcal{D}',\mathcal{P}' \vdash Q' : \mathcal{D}'',\mathcal{P}''
  }
  {\qcontext{Q\;Q'}{\mathcal{D}'',\mathcal{P}''}}
\end{eqnarray}
\caption{GIPL Semantics}
\label{fig:gipl-semantics}
\label{fig:ilucid-semantics}
\end{footnotesize}
\end{figure}

\begin{figure}[htb!]
\footnotesize
\begin{eqnarray}
{\mathbf{E_{E.did}}} &:& \frac
  {\johndef(E.\myid)=(\texttt{dim})}
  {\context{E.\myid}{\myid.\myid}}\label{sem:gipl:e.e.did}
\end{eqnarray}
\normalsize
\caption{Higher-Order Context Dot Operator}
\label{fig:hoc-dot-operator}
\end{figure}

In the implementing system, {\gipsy}, the {\gipl} is the generic counterpart of all the {\lucid} programming languages.
Like {\ilucid}, which it is derived from, it has only the two standard intensional operators: \verb|E @ C| for evaluating an expression \verb|E| in context \verb|C|, and \verb|#d| for determining the position in dimension \verb|d| of the current context of evaluation in the context space~\cite{paquetThesis}.
{\sipl}s are {\lucid} dialects (Specific Intensional Programming Languages) with their own attributes and objectives. Theoretically, all {\sipl}s can be translated into the {\gipl}~\cite{paquetThesis}.
All the {\sipl}s conservatively extend the {\gipl} syntactically and semantically.
The remainder of this section presents a relevant piece of {\lucx} as a conservative extension to {\gipl}.
The
semantics of {\gipl}
is
presented in
\xf{fig:gipl-semantics}.
The excerpt of
semantic rules of
{\lucx}
are then presented as a conservative extension to
{\gipl}
in
\xf{fig:lucx-semantics}.
Following is the description of the {\gipl} semantic rules as presented in~\cite{paquetThesis}:
\[\mathcal{D} \vdash E : v\]
%
tells that under the {\it{definition environment}}~$\mathcal{D}$, expression~$E$
would evaluate to value~$v$.
\[\context{E}{v}\]
%
specifies that in the definition environment~$\mathcal{D}$, and in
the {\it{evaluation context}}~$\mathcal{P}$~(sometimes also referred to as a
{\em{point}} in the context space), expression~$E$ evaluates to~$v$.
The definition environment~$\mathcal{D}$ retains the definitions of
all of the identifiers that appear in a {\lucid} program, as created with
the semantic rules 13-16 in \xf{fig:gipl-semantics}.  It is
therefore a partial function
\[ \mathcal{D} : \mathbf{Id} \rightarrow \mathbf{IdEntry} \]
where $\mathbf{Id}$ is the set of all possible identifiers and
$\mathbf{IdEntry}$, 
has five possible
kinds of value, one for each of the kinds of identifier:
%
\begin{inparaenum}
\item
  \emph{Dimensions} define the coordinate pairs, in which one can navigate
  with the $\Lhash$ and $\LSat$ operators.  Their $\mathbf{IdEntry}$ is
  simply $(\mathtt{dim})$.
\item
  \emph{Constants} are external entities that provide a single value,
  regardless of the context of evaluation. Examples are integers and Boolean values.  Their
  $\mathbf{IdEntry}$ is $(\mathtt{const},c)$, where $c$~is the value
  of the constant.
\item
  \emph{Data operators} are external entities that provide memoryless
  functions. Examples are the arithmetic and Boolean functions.  The
  constants and data operators are said to define the \emph{basic
  algebra} of the language.  Their $\mathbf{IdEntry}$ is
  $(\mathtt{op},f)$, where $f$~is the function itself.
\item
  \emph{Variables} carry the multidimensional streams.  Their
  $\mathbf{IdEntry}$ is $(\mathtt{var},E)$, where $E$~is the {\lucid}
  expression defining the variable. It should be noted that this
  semantics makes the assumption that all variable names are
  unique. This constraint is easy to overcome by performing
  compile-time renaming or using a nesting level environment scope
  when needed.
\item
  \emph{Functions} are non-recursive {\gipl} user-defined functions.  Their
  $\mathbf{IdEntry}$ is $(\mathtt{func},id_i,E)$, where the $id_i$ are
  the formal parameters to the function and $E$~is the body of the
  function.
  In this paper we do not discuss the semantics of recursive functions.
\end{inparaenum}

The evaluation context~$\mathcal{P}$, which is changed when the
$\LSat$ operator is evaluated, or a dimension is declared in a \api{where} clause, associates
a {\it{tag}} (i.e. an index) to each relevant dimension. It is, therefore, a partial function
\[ \mathcal{P} : \mathbf{Id} \rightarrow \mathbf{N} \]
Each type of identifiers can only be used in the appropriate situations. Identifiers of type \texttt{op}, \texttt{func}, and \texttt{dim}
evaluate to themselves~(\xf{fig:gipl-semantics}, rules \ref{sem:gipl:e.opid},\ref{sem:gipl:e.did},\ref{sem:gipl:e.fid}). Constant identifiers (\texttt{const}) evaluate to the corresponding constant~(\xf{fig:gipl-semantics}, rule \ref{sem:gipl:e.cid}). Function calls, resolved by the $\mathbf{E_{fct}}$ rule~(\xf{fig:gipl-semantics}, rule \ref{sem:gipl:e.fct}), require the renaming of the formal parameters into the actual parameters (as represented by $E'[\myid_i\leftarrow E_i]$). The function $\mathcal{P}'=\mathcal{P}\mydagger[\myid\mapsto v'']$ specifies that $\mathcal{P}'(x)$ is $v''$ if $x=\myid$, and $\mathcal{P}(x)$ otherwise. The rule for the \api{where} clause, $\mathbf{E_w}$~(\xf{fig:gipl-semantics}, rule \ref{sem:gipl:e.w}), which corresponds to the syntactic expression $E\;\mathtt{where}\;Q$,
evaluates $E$ using the definitions $Q$ therein.
The additions to the definition environment $\mathcal{D}$ and context of evaluation $\mathcal{P}$ made by the $\mathbf{Q}$ rules~(\xf{fig:gipl-semantics}, rules~\ref{sem:gipl:q.dim},\ref{sem:gipl:q.id},\ref{sem:gipl:q.fid}) are local to the current \api{where} clause.
This is represented by the fact that the $\mathbf{E_w}$ rule returns neither $\mathcal{D}$ nor $\mathcal{P}$.
The $\mathbf{Q_{dim}}$ rule adds a dimension to the definition environment and, as a convention, adds this dimension to the context of evaluation with tag~$0$~(\xf{fig:gipl-semantics}, rule~\ref{sem:gipl:q.dim}). The $\mathbf{Q_{id}}$ and $\mathbf{Q_{fid}}$ simply add variable and function identifiers along with their definition to the definition environment~(\xf{fig:gipl-semantics}, rules~\ref{sem:gipl:q.id},\ref{sem:gipl:q.fid}).

As a conservative extension to {\gipl}, {\lucx}'s semantics introduces the notion of {\it{context}} as a building block into the semantic rules, i.e. {\it{context as a first-class value}}, as described by the rules in~\xf{fig:lucx-semantics}.
In {\lucx}, semantic rule~\ref{sem:lucx:e.construction.cxt} (\xf{fig:lucx-semantics}) creates a context as a semantic item and returns it as a context $\mathcal{P}$ that can then be used by rule~\ref{sem:lucx:e.at.cxt} to navigate to this context by making it override the current context.
{\gipl}'s semantic rule~\ref{sem:gipl:e.op} is still valid for the definition of the context operators, where the actual parameters evaluate to values $v_i$ that are contexts $\mathcal{P}_i$.
The semantic rule~\ref{sem:lucx:e.hash.cxt} expresses that the \api{\#} symbol evaluates to the current context. When used as a parameter to the context calculus operators, this allows for the generation of contexts relative to the current context of evaluation.

\begin{figure}[htb!]
\begin{footnotesize}
\begin{eqnarray}
{\mathbf{E_{\#(cxt)}}} &:& \frac
  {
  }%
  {\context{\#}{\mathcal{P}}}\label{sem:lucx:e.hash.cxt}\\\nonumber\\
{\mathbf{E_{construction(cxt)}}} &\!\!\!\!\!\!\!\!\!\!:\!\!\!\!\!\!\!\!\!\!& \frac
   {
    \begin{array}{l}
    \context{E_{d_{j}}}{\myid_{j}}\qquad
    \johndef(\myid_{j})=(\texttt{dim})\\
    \context{E_{i_{j}}}{v_{j}}\qquad
    \mathcal{P}' = \mathcal{P}_{0}\mydagger[\myid_{1}\mapsto v_{1}]\mydagger \ldots \mydagger [\myid_{n}\mapsto v_{n}]
    \end{array}
   }
   {
    \context{[E_{d_{1}}:E_{i_{1}},E_{d_{2}}:E_{i_{2}},\ldots,E_{d_{n}}:E_{i_{n}}]}{\mathcal{P}'}
   }\label{sem:lucx:e.construction.cxt}\\\nonumber\\
{\mathbf{E_{at(cxt)}}} &:& \frac
  {\context{E'}{\mathcal{P'}}\qquad
  \mathcal{D},\mathcal{P}\mydagger{\mathcal{P'}} \vdash E : v
  }
  {\context{E\;@\;E'}{v}}\label{sem:lucx:e.at.cxt}
\end{eqnarray}
\caption{Conservative Semantic Rules Introduced by {\lucx}}
\label{fig:lucx-semantics}
\end{footnotesize}
\end{figure}

\section{Conclusion}

While the list of Isabelle's proofs is incomplete at the time of the writing of this
manuscript some formalization in Isabelle took place, and the work on them is
currently on-going.

\subsection{Results}

Due to a non-standard nature of the {\lucid} language (as opposed to standard imperative languages),
it takes some time to understand the full scope of some of its details and model them.
This complicates a way to model its operators, expressions, overall meaning in Isabelle.
This fact resulted in several trials and attempts to approach the language, from fairly
complex
to fairly basic -- plain integers
and pipelined processing and basic index support.
They are not fully complete, but some of the basic properties
are modeled and proven; please refer to the Isabelle sources for details (once completed
it is planned to be released as a part of the Archive of Formal Proofs at~\cite{afp}).

\begin{itemize}
\item
	The \texttt{IntegerLucid} Isabelle file is the most
	developed out of all as far as definition and exploitation of intensional
	operators of classical Lucid concerned. It is called ``integer'' because all the
	streams and dimensions and all operators around them play with integers,
	natural numbers, and in rarer cases Booleans.
	There are no identifiers in there.
	The Isabelle file contains three theories: \texttt{OriginalLucidOperators},
	\texttt{LucidOperators}, and \texttt{IntegerLucid}. The first
	models classical Lucid operators\index{Lucid!operators} as pipelined dataflows. The second adds up
	some index support and proves equivalence to the first definitions. The latter
	provides new definitions of the intensional operators through \api{@} and \api{\#},
	defines meaning functions, propositions, and lemmas from~\cite{paquetThesis}.
	Integer Lucid proves the example for $N$ @.d $2 = 44$ for the \texttt{at()}.

\item
	The \texttt{BasicLucid} theory\index{Lucid!BasicLucid} is currently
	the second one derived to support Lucid definitions. It is an extension of IntegerLucid by adding
	identifiers.
	\olucidop{asa} and \olucidop{upon} are in this theory.

\item
	The \texttt{LucidSemanticRules} theory is meant to have the meaning of complete semantic rules
	and
	proven, but it only has a definition of a Hoare tuple~\cite{moellerhoare} and a meaning
	function for it.

\item
	The \texttt{CommonLucidTypes} theory is used by all (most) theories and defines
	some common types used by most~\cite{gipsy-type-system-lpar08}.

\item
\file{ForensicLucid.thy},
\file{GIPL.thy},
\file{IndexicalLucid.thy},
\file{JLucid.thy},
\file{JOOIP.thy},
\file{Lucx.thy},
\file{ObjectiveLucid.thy}
are the theories under current development with some results from the above.
The completed work will have a complete list of the files publicly available
and submitted to the AfP~\cite{afp}.

\end{itemize}

%
%

\subsection{Future Work}
\label{sect:future-work}

The near-future work will consist primarily of the following items:

\begin{itemize}
\item
Complete semantics of all the mentioned Lucid dialects and their formalization with Isabelle.

\item
Augment the language specification to include the Depmster-Shafer theory~\cite{shafer-evidence-theory,prob-argumentation-systems}
of evidence to allow weights for claims, credibility, belief, and plausibility
parameters.

\item
Prove semantic rules involving intensional data warehouse.

\item
Implementation of the {\flucid} compiler, run-time and interactive development environments.

\end{itemize}

\section{Acknowledgments}

This research and development work was funded in part by NSERC and the Faculty
of Engineering and Computer Science of Concordia University,
Montreal, Canada. Thanks to Drs. Mourad Debbabi, Patrice Chalin, Peter
Grogono on valuable suggestions used in this work.

%
\label{sect:bib}
\bibliographystyle{splncs}
\bibliography{flucid-isabelle-tphols08}


%
%

\appendix
\section*{{\Huge Appendix}}

\section{Lucid Axioms, Theorems, and Proofs}
\label{appdx:lucid-proofs}

Here we present and extend the notion of the formalisms
from Paquet~\cite{paquetThesis} and extend them on to
the present work.

\subsection{Streaming and Basic Operators}
\label{lucid:pipeline}

The origins of {\lucid} date back to 1974.  At that time,
Ashcroft\index{Ashcroft, E.~A.} and Wadge\index{Wadge, W.~W.} were working on
a purely declarative language, in which iterative algorithms could be
expressed naturally, which eventually resulted in~\cite{nonprocedural-iterative-lucid-77}.
Their work fits into the broad area
of research into program semantics and verification.  It would later turn
out that their work is also relevant to the
dataflow networks and coroutines of Kahn\index{Kahn, G.} and
MacQueen\index{MacQueen, D.~B.}~\cite{ka74,ka77}. In the original
{\lucid}~(whose operators are in \olucidop{this font}),
streams\index{stream} were defined in a
pipelined\index{dataflow!pipelined} manner, with two separate
definitions: one for the initial element, and another one for the
subsequent elements.  For example, the equations

\[
\begin{array}{lcl}
\molucidop{first} X & = & 0\\
\molucidop{next} X  & = & X+1\\
\end{array}
\]

\noindent define variable $X$ to be a stream, such that
\begin{eqnarray*}
x_0 & = & 0\\
x_{i+1} & = & x_i + 1
\end{eqnarray*}

\noindent In other words,
\[
\begin{array}{lcl}
0 & = & (0, 0, 0, ..., 0, ...)\\
X & = & (x_0, x_1, \ldots, x_i, \ldots) = (0,1,\ldots,i,\ldots)
\end{array}
\]
Similarly, the equations

\[
\begin{array}{lcl}
\molucidop{first} X & = & X\\
\molucidop{next} Y & = & Y + \molucidop{next} X\\
\end{array}
\]

\noindent define variable $Y$ to be the running sum of $X$, i.e.
\begin{eqnarray*}
y_0 & = & x_0\\
y_{i+1} & = & y_i + x_{i+1}
\end{eqnarray*}

\noindent In other words,

\[
Y= (y_0, y_1, \ldots, y_i, \ldots) =
\left(0,1,\ldots,{\textstyle i(i+1)\over2},\ldots\right)
\]

\noindent
It soon became clear that a ``new'' operator at the time, \Tfby\,
\index{followed@{\em{followed by}}}({\em{followed by}}) can be
used to define such typical situations.  Hence, the above two variables
could be defined as follows:

\[
\begin{array}{lcl}
X & = & 0 \molucidop{fby} X + 1\\
Y & = & X \molucidop{fby} Y + \molucidop{next} X\\
\end{array}
\]

\noindent
As a result, we can summarize the three basic operators of the original
{\lucid}.

\sdefinition
{
If $X = (x_0, x_1, \ldots, x_i, \ldots)$ and
$Y = (y_0, y_1, \ldots, y_i, \ldots)$, then
\label{pipeline}
\[
\begin{array}{llcl}
(1) & \molucidop{first} X  & \eqdef & (x_0, x_0, \ldots, x_0, \ldots)\\
(2) & \molucidop{next} X   & \eqdef & (x_1, x_2, \ldots, x_{i+1}, \ldots)\\
(3) & X \molucidop{fby} Y & \eqdef & (x_0, y_0, y_1, \ldots, y_{i-1}, \ldots)\\
\end{array}
\]
}

\noindent
Here parallels can be drawn to the list operations, where
\Tfirst\ corresponds to~\texttt{head},
\Tnext\ corresponds to~\texttt{tail}, and
\Tfby\ corresponds to~\texttt{cons}.
When these operators are combined with \index{Landin, P.~J.}Landin's
\index{ISWIM}ISWIM~\cite{ISWIM} ({\em If You See What I Mean}),
essentially typed $\lambda$-calculus with
syntactic sugar, it becomes possible to define complete {\lucid}
programs.  The following three derived operators have turned out to be
very useful (we will use them later in the text):

\sdefinition
{
\label{initdefs}
\[
\begin{array}{llcl}
(1) & X \molucidop{wvr} Y  & \eqdef &
\begin{array}[t]{@{}l@{}l@{}l}
    \Lif \molucidop{first} Y &\Lthen X \molucidop{fby} &(\molucidop{next} X \molucidop{wvr} \molucidop{next} Y)\\
           &\Lelse          &(\molucidop{next} X \molucidop{wvr} \molucidop{next} Y)
\end{array}\\
(2) & X \molucidop{asa} Y  & \eqdef & \molucidop{first} (X \molucidop{wvr} Y) \\
(3) & X \molucidop{upon} Y & \eqdef &
\begin{array}[t]{@{}l@{}l@{}l}
  X \molucidop{fby} (\Lif \molucidop{first} Y
            &\Lthen (\molucidop{next} &X \molucidop{upon} \molucidop{next} Y)\\
            &\Lelse (        &X \molucidop{upon} \molucidop{next} Y))\\
\end{array}
\end{array}
\]
}

\noindent
Where \Twvr\ stands for {\em{whenever}}\index{whenever@{\em{whenever}}},
\Tasa\ stands for {\em{as soon as}}\index{as@{\em{as soon as}}} and
\Tupon\ stands for \index{advances@{\em{advances upon}}}{\em{advances upon}}.

\subsection{Random Access to Streams}
\label{taggedtoken}

With the original {\lucid} operators, one could only define programs
with pipelined dataflows\index{dataflow!pipeline}, i.e. in which the
$(i+1)$-th element in a stream is only computed once the $i$-th
element has been computed.  This situation is potentially wasteful of
resources, since the $i$-th element might not necessarily be required.
More importantly, it only allows sequential access into streams.

By taking a different approach, it is possible to have random access
into streams, using an index~$\Lhash$ corresponding to the current
position, the current context of evaluation.  No longer are we manipulating infinite extensions
(streams)\index{stream}, rather we are defining computation according
to a context\index{context} (here a single integer).  We have set out
on the road to intensional programming\index{intensional!programming}.
We redefine all \olucidop{original} Lucid operators in terms
of the operators $\Lhash$ and~$\LSat$:

\sdefinition
{
\label{athash}
\[
\begin{array}{llcl}
(1) & \Lhash 		& \eqdef & 0 \molucidop{fby} (\Lhash + 1)\\
(2) & X \Lat Y  	& \eqdef & \Lif Y=0 \Lthen \molucidop{first} X \\
    & 			&   	 & \Lelse (\molucidop{next} X) \Lat (Y-1)\\
\end{array}
\]
}

\clearpage

\noindent
Further, we give definitions for the original operators
using these two baseline operators. In so doing, we will use the
following axioms.

\saxiom
{
\label{axioms}
Let $i\geq 0$.
\[
\begin{array}{ll}
(1) & [c]_i = c\\
(2) & [X+c]_i = [X]_i+c\\
(3) & [\molucidop{first} X]_i = [X]_0\\
(4) & [\molucidop{next} X]_i = [X]_{i+1}\\
(5) & [X\molucidop{fby} Y]_0 = [X]_0\\
(6) & [X\molucidop{fby} Y]_{i+1} = [Y]_i\\
(7) & \Lif \texttt{true} \Lthen [X]_i \Lelse [Y]_i = [X]_i\\
(8) & \Lif \texttt{false} \Lthen [X]_i \Lelse [Y]_i = [Y]_i\\
(9) & [\Lif C \Lthen X \Lelse Y]_i =
\Lif [C]_i \Lthen [X]_i \Lelse [Y]_i\\
\end{array}
\]
}

\noindent
Prior giving the re-definitions of the standard {\lucid} operators,
we show some basic properties of \LSat\ and \Lhash.  We will use
throughout the discussion here $[X]_i$ instead of $x_i$, as it allows
for greater readability. Furthermore, we will, as is standard,
write $X = Y$ whenever we have
\[(\forall i : i\geq 0 : [X]_i=[Y]_i)\]

\sproposition
{
\label{prophash}
Let $i\geq 0$.
\[
\begin{array}{ll}
(1) & [\Lhash]_i = i\\
(2) & [X \Lat Y]_i = [X]_{[Y]_i}\\
\end{array}
\]
}

\paragraph{Proof} \mbox{}\\
(1) Proof by induction over~$i$.

\textbf{Base step} ($i=0$).

\[
\begin{array}{lclp{8.1cm}l}
[\Lhash]_0  & = & [0\molucidop{fby} (\Lhash+1)]_0   &&  \mbox{Defn.~\ref{athash}.1}\\
                 & = & [0]_0   &&  \mbox{Axiom~\ref{axioms}.5}\\
                 & = & 0       &&  \mbox{Axiom~\ref{axioms}.1}\\
\end{array}
\]

\textbf{Induction step} ($i=k+1$).
Suppose $(\forall i : i\leq k : [\Lhash]_i = i)$.

\[
\begin{array}{lclp{7.35cm}l}
[\Lhash]_{k+1} & = & [0\molucidop{fby} (\Lhash+1)]_{k+1} && \mbox{Defn.~\ref{athash}.1}\\
               & = & [\Lhash+1]_k                        && \mbox{Axiom~\ref{axioms}.6}\\
               & = & [\Lhash]_k+1                        && \mbox{Axiom~\ref{axioms}.2}\\
               & = & k+1                                 && \mbox{Ind.\ Hyp.}\\
\end{array}
\]

\noindent
Hence $(\forall i : i\geq 0 : [\Lhash]_i = i)$.
\medskip

\noindent
{(2)} Let $i\geq 0$.  We will prove by induction over~$y_i$ that
$y_i\geq 0 \Rightarrow [X\Lat Y]_i = [X]_{[Y]_i}$.

\textbf{Base step} ($y_i=0$).

\[
\begin{array}{lclp{0cm}l}
[X \Lat Y]_i
&=& [\Lif Y=0 \Lthen \molucidop{first} X \Lelse (\molucidop{next} X) \Lat (Y-1)]_i   &&  \mbox{Defn.~\ref{athash}.2}\\
&=& \Lif [Y=0]_i \Lthen [\molucidop{first} X]_i \Lelse [(\molucidop{next} X) \Lat (Y-1)]_i &&  \mbox{Axiom~\ref{axioms}.9}\\
&=& \Lif [Y]_i=0 \Lthen [\molucidop{first} X]_i \Lelse [(\molucidop{next} X) \Lat (Y-1)]_i && \mbox{Axiom~\ref{axioms}.2}\\
&=& [\molucidop{first} X]_i && \mbox{Axiom~\ref{axioms}.7}\\
&=& [X]_0 && \mbox{Axiom~\ref{axioms}.3}\\
&=& [X]_{[Y]_i} && \mbox{Hypothesis}\\
\end{array}
\]

\textbf{Induction step} ($y_i=k+1$).
Suppose $(\forall i : i\leq k : [\Lhash]_i = i)$.

\[
\begin{array}{lclp{0cm}l}
[X \Lat Y]_i
&=& [\Lif Y=0 \Lthen \molucidop{first} X \Lelse (\molucidop{next} X) \Lat (Y-1)]_i   &&  \mbox{Defn.~\ref{athash}.2}\\
&=& \Lif [Y=0]_i \Lthen [\molucidop{first} X]_i \Lelse [(\molucidop{next} X) \Lat (Y-1)]_i &&  \mbox{Axiom~\ref{axioms}.9}\\
&=& \Lif [Y]_i=0 \Lthen [\molucidop{first} X]_i \Lelse [(\molucidop{next} X) \Lat (Y-1)]_i && \mbox{Axiom~\ref{axioms}.2}\\
&=& [(\molucidop{next} X) \Lat (Y-1)]_i &&  \mbox{Axiom~\ref{axioms}.8}\\
&=& [\molucidop{next} X]_{[Y-1]_i} &&  \mbox{Ind. Hyp.}\\
&=& [\molucidop{next} X]_{[Y]_i-1} &&  \mbox{Axiom~\ref{axioms}.2}\\
&=& [X]_{[Y]_i-1+1} &&  \mbox{Axiom~\ref{axioms}.4}\\
&=& [X]_{[Y]_i} && \mbox{Arith.}\\
\end{array}
\]

\noindent
Hence $(\forall i : i\geq 0 :
      [Y]_i\geq 0 \Rightarrow ([X \Lat Y]_i = [X]_{[Y]_i}))$.
\hfill $\Box$


\sdefinition
{
\label{redefs}
\[
\begin{array}{llcl}
(1) & \mfirst X  	& \eqdef & X \Lat 0\\
(2) & \mnext X   	& \eqdef & X \Lat (\Lhash+1)\\
(3) & X \mfby Y & \eqdef &
    \Lif \Lhash=0 \Lthen X \Lelse Y \Lat (\Lhash-1)\\
(4) & X \mwvr Y & \eqdef & X \Lat T\\
    &&& \texttt{where}\\
(4.1)  &&& \quad T = U \mfby U \Lat (T+1)\\
(4.2)  &&& \quad U = \Lif Y \Lthen \Lhash \Lelse \mnext U\\
       &&& \texttt{end}\\
(5) & X \masa Y & \eqdef & \mfirst(X \mwvr Y)\\
(6) & X \mupon Y & \eqdef & X \Lat W\\
    &&& \texttt{where}\\
(6.1)  &&& \quad W = 0 \mfby \Lif Y \Lthen (W+1) \Lelse W\\
       &&& \texttt{end}\\
\end{array}
\]
}

\noindent
The advantage of these new definitions is that they do not use any
form of recursive function definitions.  Rather, all of the
definitions are iterative, and in practice, more easily implemented in
an efficient manner.  We prove below that the \lucidop{new}
definitions are equivalent to the \olucidop{old} ones.

\sproposition
{
\label{firstprop}
$\mfirst X = \molucidop{first} X$.
}

\paragraph{Proof}
Let $i\geq 0$.  Then

\[
\begin{array}{lclp{7.65cm}l}
[\mfirst X]_i  & = & [X\Lat 0]_i   &&  \mbox{Defn.~\ref{redefs}.1}\\
                 & = & [X]_{[0]_i}   &&  \mbox{Prop.~\ref{prophash}.2}\\
                 & = & [X]_0         &&  \mbox{Axiom~\ref{axioms}.1}\\
	 	 & = & [\molucidop{first} X]_i &&  \mbox{Axiom~\ref{axioms}.3}\\
\end{array}
\]

\noindent
Hence $\mfirst X = \molucidop{first} X$. \hfill $\Box$

\clearpage

\sproposition
{
$\mnext  X = \molucidop{next} X$.
}

\paragraph{Proof}
Let $i\geq 0$.  Then

\[
\begin{array}{lclp{7.3cm}l}
[\mnext X]_i & = & \bigl[X\Lat (\Lhash+1)\bigr]_i
   &&  \mbox{Defn.~\ref{redefs}.2}\\
& = & [X]_{[\Lhash+1]_i}                   &&  \mbox{Prop.~\ref{prophash}.2}\\
& = & [X]_{[\Lhash]_i+1}                   &&  \mbox{Axiom~\ref{axioms}.2}\\
& = & [X]_{i+1}                            &&  \mbox{Prop.~\ref{prophash}.1}\\
& = & [\molucidop{next} X]_i               &&  \mbox{Axiom~\ref{axioms}.4}\\
\end{array}
\]

\noindent
Hence $\mnext X = \molucidop{next} X$. \hfill $\Box$

\sproposition
{
$X \mfby Y = X \molucidop{fby} Y$.
}

\paragraph{Proof} Proof by induction over~$i$.

\textbf{Base step} ($i=0$).

\[
\begin{array}{lclp{2.1cm}l}
[X \mfby Y]_0 & = &
[\Lif \Lhash=0 \Lthen X \Lelse Y \Lat (\Lhash-1)]_0
                                                && \mbox{Defn.~\ref{redefs}.3}\\
&=&\Lif [\Lhash=0]_0 \Lthen [X]_0 \Lelse [Y \Lat (\Lhash-1)]_0
                                                && \mbox{Axiom~\ref{axioms}.9}\\
&=&\Lif [\Lhash]_0=0 \Lthen [X]_0 \Lelse [Y \Lat (\Lhash-1)]_0
                                                && \mbox{Defn.~\ref{axioms}.2}\\
&=&\Lif 0=0 \Lthen [X]_0 \Lelse [Y \Lat (\Lhash-1)]_0
                                              && \mbox{Prop.~\ref{prophash}.1}\\
& = & [X]_0					&& \mbox{Axiom~\ref{axioms}.7}\\
& = & [X \molucidop{fby} Y]_0				&& \mbox{Axiom~\ref{axioms}.5}\\
\end{array}
\]

\textbf{Induction step} ($i=k+1$).

\[
\begin{array}{lclp{0.5cm}l}
[X \mfby Y]_{k+1} & = &
\bigl[\Lif \Lhash=0 \Lthen X \Lelse Y \Lat (\Lhash-1)\bigr]_{k+1}
                                           && \mbox{Defn.~\ref{redefs}.3}\\
&=&\Lif [\Lhash=0]_{k+1} \Lthen [X]_{k+1} \Lelse [Y \Lat (\Lhash-1)]_{k+1}
                                                && \mbox{Axiom~\ref{axioms}.9}\\
&=&\Lif [\Lhash]_{k+1}=0 \Lthen [X]_{k+1} \Lelse [Y \Lat (\Lhash-1)]_{k+1}
                                                && \mbox{Axiom~\ref{axioms}.1}\\
&=&\Lif k+1=0 \Lthen [X]_{k+1} \Lelse [Y \Lat (\Lhash-1)]_{k+1}
                                             && \mbox{Prop.~\ref{prophash}.1}\\
& = & \bigl[Y \Lat (\Lhash-1)\bigr]_{k+1} && \mbox{Axiom~\ref{axioms}.8}\\
& = & \bigl[Y\bigr]_{[\Lhash-1]_{k+1}} && \mbox{Prop.~\ref{prophash}.2}\\
& = & \bigl[Y\bigr]_{[\Lhash]_{k+1}-1} && \mbox{Axiom~\ref{axioms}.2}\\
& = & \bigl[Y\bigr]_k			&& \mbox{Prop.~\ref{prophash}.1}\\
& = & [X \molucidop{fby} Y]_{k+1}		&& \mbox{Axiom~\ref{axioms}.6}\\
\end{array}
\]

\noindent
Hence $(\forall i : i\geq 0 : [X \mfby Y]_i = [X \molucidop{fby} Y]_i)$.
Hence $\mfby = \molucidop{fby}$. \hfill $\Box$

\noindent
The proof for \Twvr\ is more complicated, as it requires relating an
iterative definition to a recursive definition.  We will therefore
need four lemmas that refer to variables~$T$ and~$U$
in the text in Definitions~\ref{redefs}.4.1 and~\ref{redefs}.4.2.
In addition, we must define the \emph{rank} of a Boolean stream.
Finally, we will have to introduce another set of axioms, that
allow us to compare two entire streams, as opposed to particular elements
in the two streams.

\clearpage

\saxiom
{
\label{equivaxioms}
Let $i\geq 0$.
\[
\begin{array}{ll}
(1) & X^0 = X\\
(2) & [X^i]_0 = [X]_i\\
(3) & \molucidop{first} X^i = [X]_i\\
(4) & \molucidop{next} X^i = X^{i+1}\\
(5) & \molucidop{next} (X \molucidop{fby} Y) = Y\\
(6) & (\molucidop{first} X) \molucidop{fby} Y = X \molucidop{fby} Y\\
(7) & \Lif \texttt{true} \Lthen X \Lelse Y = X\\
(8) & \Lif \texttt{false} \Lthen X \Lelse Y = Y\\
\end{array}
\]
}

\sdefinition
{
\label{rankdefn}
Let $Y$ be a Boolean stream.
\[
\begin{array}{llcl}
(1) & \texttt{rank}(-1,Y) & \eqdef & -1\\
(2) & \texttt{rank}(i+1,Y) & \eqdef &
  \mathrm{min}\{k : k> \texttt{rank}(i,Y) : [Y]_k=\texttt{true}\}\\
\end{array}
\]
}

\noindent
Further, we write $r_i$ for $\texttt{rank}(i,Y)$.

\slemma
{
\label{wvrlemma}
$(\forall i : i\geq -1 :
  (\forall j : r_i < j \leq r_{i+1} :
    X^{j} \molucidop{wvr} Y^{j} = X^{r_{i+1}} \molucidop{wvr} Y^{r_{i+1}}))$.
}

\paragraph{Proof}
Let $i\geq -1$. Proof by downwards induction over $j$. Note that $r_i<r_{i+1}$.

\textbf{Base step} ($j=r_{i+1}$).
\[
\begin{array}{lclp{5.9cm}l}
X^{r_{i+1}} \molucidop{wvr} Y^{r_{i+1}} & = & X^{r_{i+1}} \molucidop{wvr} Y^{r_{i+1}}
                      && \mbox{Identity}\\
\end{array}
\]

\textbf{Induction step} ($j=k-1$, $j>r_i$).

\[
\begin{array}{lclp{0.9cm}l}
X^{k-1} \molucidop{wvr} Y^{k-1} & = & \Lif \molucidop{first} Y^{k-1}\begin{array}[t]{@{}l}
                     \Lthen X^{k-1} \molucidop{fby} X^k \molucidop{wvr} Y^k\\
                     \Lelse X^k \molucidop{wvr} Y^k\\
                     \end{array}&&\mbox{Defn.\ \ref{initdefs}.1}\\
               & = & \Lif [Y]_{k-1}\begin{array}[t]{@{}l}
                     \Lthen X^{k-1} \molucidop{fby} X^k \molucidop{wvr} Y^k\\
                     \Lelse X^k \molucidop{wvr} Y^k\\
                     \end{array}&&\mbox{Axiom \ref{equivaxioms}.3}\\
               & = & X^k \molucidop{wvr} Y^k
                     &&\mbox{Axiom \ref{equivaxioms}.8}\\
               & = & X^{r_{i+1}} \molucidop{wvr} Y^{r_{i+1}}
                     &&\mbox{Ind.\ Hyp.}\\
\end{array}
\]

\noindent
Hence,
$(\forall i : i\geq -1 :
  (\forall j : r_i < j \leq r_{i+1} :
    X^{j} \molucidop{wvr} Y^{j} = X^{r_{i+1}} \molucidop{wvr} Y^{r_{i+1}}))$.
\hfill $\Box$

\slemma
{
\label{wvrlemma1}
$(\forall i : i\geq 0 : (X \molucidop{wvr} Y)^i = X^{r_i} \molucidop{wvr} Y^{r_i})$.
}

\paragraph{Proof} Proof by induction over $i$.

\textbf{Base step} ($i=0$).

\[
\begin{array}{lclp{7.2cm}l}
(X \molucidop{wvr} Y)^0 &=& X \molucidop{wvr} Y&&\mbox{Axiom~\ref{equivaxioms}.1}\\
               &=& X^0 \molucidop{wvr} Y^0&&\mbox{Axiom~\ref{equivaxioms}.1}\\
               &=& X^{r_0} \molucidop{wvr} Y^{r_0} && \mbox{Lemma~\ref{wvrlemma}}\\
\end{array}
\]

\textbf{Induction step} ($i=k+1$).

\[
\begin{array}{lcl@{}p{0.0cm}l}
(X \molucidop{wvr} Y)^{k+1} &=& \molucidop{next}((X \molucidop{wvr} Y)^k)
&&\mbox{Axiom \ref{equivaxioms}.4}\\
                   &=& \molucidop{next}(X^{r_k} \molucidop{wvr} Y^{r_k})
&&\mbox{Ind.\ Hyp.}\\
&=& \molucidop{next}(\Lif \molucidop{first} Y^{r_k}\begin{array}[t]{@{}l}
            \Lthen X^{r_k} \molucidop{fby} X^{{r_k}+1} \molucidop{wvr} Y^{{r_k}+1}\\
            \Lelse X^{{r_k}+1} \molucidop{wvr} Y^{{r_k}+1})\\
            \end{array}
&&\mbox{Defn.\ \ref{initdefs}.1}\\
&=& \molucidop{next}(\Lif [Y]_{r_k} \begin{array}[t]{@{}l}
            \Lthen X^{r_k} \molucidop{fby} X^{{r_k}+1} \molucidop{wvr} Y^{{r_k}+1}\\
            \Lelse X^{{r_k}+1} \molucidop{wvr} Y^{{r_k}+1})\\
            \end{array}
&&\mbox{Axiom \ref{equivaxioms}.3}\\
&=& \molucidop{next}(X^{r_k} \molucidop{fby} X^{{r_k}+1} \molucidop{wvr} Y^{{r_k}+1})
&&\mbox{Axiom \ref{equivaxioms}.7}\\
&=& X^{{r_k}+1} \molucidop{wvr} Y^{{r_k}+1}
&&\mbox{Axiom \ref{equivaxioms}.5}\\
&=& X^{r_{k+1}} \molucidop{wvr} Y^{r_{k+1}}
&&\mbox{Lemma~\ref{wvrlemma}}\\
\end{array}
\]

\noindent
Hence, $(\forall i : i\geq 0 : (X \molucidop{wvr} Y)^i = X^{r_i} \molucidop{wvr} Y^{r_i})$.
\hfill $\Box$

\slemma
{
\label{wvrlemma2}
$(\forall i : i\geq -1 :
  (\forall j : r_i < j \leq r_{i+1} :
   [U]_j = r_{i+1}))$.
}

\paragraph{Proof}
Let $i\geq -1$. Proof by downwards induction over $j$. Note that $r_i<r_{i+1}$.

\textbf{Base step} ($j=r_{i+1}$).

\[
\begin{array}{lclp{3.2cm}l}
[U]_{r_{i+1}}
&=& [\Lif Y \Lthen \Lhash \Lelse \mnext U]_{r_{i+1}}
    && \mbox{Defn.~\ref{redefs}.4.2}\\
&=& \Lif [Y]_{r_{i+1}} \Lthen [\Lhash]_{r_{i+1}}
                       \Lelse [\mnext U]_{r_{i+1}}
    && \mbox{Axiom~\ref{axioms}.9}\\
&=& [\Lhash]_{r_{i+1}} && \mbox{Axiom~\ref{axioms}.7}\\
&=& r_{i+1} && \mbox{Prop.~\ref{prophash}.1}\\
\end{array}
\]

\textbf{Induction step} ($j=k-1$, $j>r_i$).

\[
\begin{array}{lclp{3.3cm}l}
[U]_{k-1} & = & [\Lif Y \Lthen \Lhash \Lelse \mnext U]_{k-1}
  && \mbox{Defn.~\ref{redefs}.4.2}\\
      & = & \Lif [Y]_{k-1} \Lthen [\Lhash]_{k-1} \Lelse [\mnext U]_{k-1}
  && \mbox{Axiom~\ref{axioms}.9}\\
      & = & [\mnext U]_{k-1} && \mbox{Axiom~\ref{axioms}.8}\\
      & = & [U]_k && \mbox{Axiom~\ref{axioms}.4}\\
      &=& r_{i+1} && \mbox{Ind.\ Hyp.}\\
\end{array}
\]

\noindent
Hence,
$(\forall i: i\geq -1 :(\forall j : r_{i-1} < j < r_i : [U]_j = r_{i+1}))$.
\hfill $\Box$

\clearpage

\slemma
{
\label{wvrlemma3}
$(\forall i : i\geq 0 : [T]_i = r_i)$.
}

\paragraph{Proof}  Proof by induction over $i$.

\textbf{Base step} ($i=0$).
\[
\begin{array}{lclp{7.1cm}l}
[T]_0 &=& [U \mfby U \Lat (T+1)]_0
    && \mbox{Defn.~\ref{redefs}.4.1}\\
&=& [U]_0
    && \mbox{Axiom~\ref{axioms}.5}\\
&=& r_0 && \mbox{Lemma~\ref{wvrlemma2}}\\
\end{array}
\]

\textbf{Induction step} ($i=k+1$).
\[
\begin{array}{lclp{6.3cm}l}
[T]_{k+1} &=& [U \mfby U \Lat (T+1)]_{k+1}
    && \mbox{Defn.~\ref{redefs}.4.1}\\
          &=& [U \Lat (T+1)]_k
    && \mbox{Axiom~\ref{axioms}.6}\\
          &=& [U]_{[T+1]_k}
    && \mbox{Prop.~\ref{prophash}.2}\\
          &=& [U]_{[T]_k+1}
    && \mbox{Axiom~\ref{axioms}.2}\\
          &=& [U]_{r_k+1}
    && \mbox{Ind.\ Hyp.}\\
          &=& r_{k+1}
    && \mbox{Lemma~\ref{wvrlemma2}}\\
\end{array}
\]

\noindent
Hence, $(\forall i : i\geq 0 : [T]_i = r_i)$.
\hfill $\Box$

\sproposition
{
\label{wvr}
$X \mwvr Y= X\molucidop{wvr} Y$.
}

\paragraph{Proof}
\mbox{}

\[
\begin{array}{lclp{1.4cm}l}
[X\mwvr Y]_i & = & [X \Lat T]_i && \mbox{Defn. 4.4}\\
	       & = & [X]_{[T]_i} && \mbox{Prop.~\ref{prophash}.2}\\
	       & = & [X]_{r_i} && \mbox{Lemma~\ref{wvrlemma3}}\\
	       & = & [X^{r_i}]_0 &&
                     \mbox{Axiom~\ref{axioms}.2}\\
	       & = & [X^{r_i}\molucidop{fby} X^{r_i+1} \molucidop{wvr} Y^{r_i+1}]_0 &&
                     \mbox{Axiom~\ref{axioms}.6}\\
      &=&  [\Lif [Y]_{r_i}
        \begin{array}[t]{@{}l}
        \Lthen X^{r_i} \molucidop{fby} X^{r_i+1} \molucidop{wvr} Y^{r_i+1}\\
        \Lelse X^{r_i+1} \molucidop{wvr} Y^{r_i+1}]_0\\
        \end{array}
&& \mbox{Axiom~\ref{equivaxioms}.7}\\
      &=&  [\Lif \molucidop{first} Y^{r_i}
        \begin{array}[t]{@{}l}
        \Lthen X^{r_i} \molucidop{fby} X^{r_i+1} \molucidop{wvr} Y^{r_i+1}\\
        \Lelse X^{r_i+1} \molucidop{wvr} Y^{r_i+1}]_0\\
        \end{array}
&& \mbox{Axiom~\ref{equivaxioms}.3}\\
	       & = & [X^{r_i}\molucidop{wvr} Y^{r_i}]_0
&& \mbox{Defn.~\ref{initdefs}.1}\\
	       & = & [(X \molucidop{wvr} Y)^i]_0 && \mbox{Lemma~\ref{wvrlemma1}}\\
	       & = & [X \molucidop{wvr} Y]_i && \mbox{Axiom~\ref{equivaxioms}.2}\\
\end{array}
\]

\noindent
Hence $X\mwvr Y = X \molucidop{wvr} Y$. \hfill $\Box$

\sproposition
{
$X \masa Y = X \molucidop{asa} Y$.
}

\paragraph{Proof}\mbox{}
\[
\begin{array}{lclp{6.8cm}l}
X \masa Y 	& = & \mfirst(X \mwvr Y)&&
   \mbox{Defn.~\ref{redefs}.5}\\
 	& = & \mfirst(X \molucidop{wvr} Y)&&
   \mbox{Prop.~\ref{wvr}}\\
 	& = & \molucidop{first}(X \molucidop{wvr} Y)&&
   \mbox{Prop.~\ref{firstprop}}\\
	& = & X \molucidop{asa} Y &&
   \mbox{Defn.~\ref{initdefs}.2}\\
\end{array}
\]

\noindent
Hence $X\masa Y = X \molucidop{asa} Y$. \hfill $\Box$

\clearpage

\slemma
{
\label{uponlemma}
$(\forall i : i\geq 0 : (X \molucidop{upon} Y)^i = X^{[W]_i} \molucidop{upon} Y^i)$
}

\paragraph{Proof}
Proof by induction over~$i$.

\textbf{Base step} ($i=0$).

\[
\begin{array}{lclp{5.8cm}l}
(X \molucidop{upon} Y)^0
  & = & X \molucidop{upon} Y
&&\mbox{Axiom~\ref{equivaxioms}.1}\\
  & = & X^0 \molucidop{upon} Y^0
&&\mbox{Axiom~\ref{equivaxioms}.1}\\
  & = & X^{[0 \mfby ...]_0} \molucidop{upon} Y^0
&&\mbox{Defn.~\ref{initdefs}.3}\\
  & = & X^{[W]_0} \molucidop{upon} Y^0
&&\mbox{Defn.~\ref{redefs}.6.1}\\
\end{array}
\]

\textbf{Induction step} ($i=k+1$).

\[
\begin{array}{lclp{3.0cm}l}
(X \molucidop{upon} Y)^{k+1}
  & = & \molucidop{next}\bigr((X \molucidop{upon} Y)^k\bigr)
&&\mbox{Axiom~\ref{equivaxioms}.4}\\
  & = & \molucidop{next}\bigl(X^{[W]_k} \molucidop{upon} Y^k\bigr)
&&\mbox{Ind.~Hyp.}\\
  & = & \Lif (\molucidop{first} Y^k)
&&\mbox{Defn.~\ref{initdefs}.3 and}\\
&&\Lthen (X^{[W]_k+1} \molucidop{upon} Y^{k+1})
&&\mbox{Axiom~\ref{equivaxioms}.5}\\
&&\Lelse (X^{[W]_k} \molucidop{upon} Y^{k+1})\\
  & = & \Lif [Y]_k
&&\mbox{Axiom~\ref{equivaxioms}.4}\\
&&\Lthen (X^{[W]_{k+1}} \molucidop{upon} Y^{k+1})
&&\mbox{Defn.~\ref{redefs}.6.1}\\
&&\Lelse (X^{[W]_k} \molucidop{upon} Y^{k+1})\\
  & = & \bigl(X^{(\Lif [Y]_k \Lthen [W]_{k+1} \Lelse [W]_k)}\bigr)
&&\mbox{Substit.}\\
&&\molucidop{upon} Y^{k+1}&&\\
  & = & X^{[W]_{k+1}} \molucidop{upon} Y^{k+1}
&&\mbox{Defn.~\ref{redefs}.6.1}\\
\end{array}
\]

\noindent
Hence,
$(\forall i : i\geq 0 : (X \molucidop{upon} Y)^i = X^{[W]_i} \molucidop{upon} Y^i)$
\hfill $\Box$

\sproposition
{
$X \mupon Y = X \molucidop{upon} Y$.
}

\paragraph{Proof}
Let $i\geq 0$.  Then

\[
\begin{array}{lclp{6.4cm}l}
[X \mupon Y]_i &=& [X \Lat W]_i
&&\mbox{Defn.~\ref{redefs}.6}\\
               &=& [X]_{[W]_i}
&&\mbox{Prop.~\ref{prophash}.2}\\
                 &=& [X^{[W]_i}]_0
&&\mbox{Axiom~\ref{equivaxioms}.2}\\
                 &=& [X^{[W]_i} \molucidop{fby} \ldots]_0
&&\mbox{Axiom~\ref{axioms}.5}\\
                 &=& [X^{[W]_i} \molucidop{upon} Y^i]_0
&&\mbox{Defn.~\ref{initdefs}.3}\\
                 &=& [X \molucidop{upon} Y]_i
&&\mbox{Lemma~\ref{uponlemma}}\\
\end{array}
\]

\noindent
Hence $X \mupon Y = X \molucidop{upon} Y$. \hfill $\Box$

\noindent
Now that the corresponding definitions are shown to be equivalent,
we can generalize and head off in the negative direction as well:

\sdefinition
{
\label{latest:fby}
\[
\begin{array}{llcl}
(1) & \mprev X  & \eqdef & X \Lat (\Lhash-1)\\
(2) & X \mfby Y & \eqdef & \Lif \Lhash \leq 0 \Lthen X
                            \Lelse Y \Lat (\Lhash-1)\\
\end{array}
\]
}

\section{Summary of the Operators' Examples}
\label{sect:new-operators-table}
\label{appdx:new-operators-table}

Here we illustrate a few basic examples of application of the {\flucid} operators (both, classical {\lucid}
and the newly introduced operators).
Assume we have two bounded (between \api{bod} and \api{eod}) streams $X$ and $Y$
of ten elements. The $X$ stream is just an ordered sequence of natural numbers
between $1$ and $10$. If queried for values below $1$ an beginning-of-data (\api{bod}) marker
would be returned; similarly if queried beyond $10$, the end-of-data marker (\api{eod}) is returned.
The $Y$ stream is a sequence of ten truth values (can be replaced with 0 for ``false'' and 1 for ``true'').
The operators applied to these streams may return bounded or unbounded streams of the same or
different length than the original depending on the definition of a particular operator.
Also assume the current dimension index is $0$. The resulting table showing the application
of the classical and the new operators is in \xt{tab:flucid-op-examples}.

\begin{table}[htb!]
\centering
\begin{tabular}{|c|c|c|c|c|c|c|c|c|c|c|c|c|c|} \hline
stream/index          &  -1 & 0   & 1  & 2  & 3  & 4  & 5  & 6  & 7  & 8  & 9   & 10  & 11  \\ \hline\hline
X                     & bod & 1   & 2  & 3  & 4  & 5  & 6  & 7  & 8  & 9  & 10  & eod & eod \\ \hline
Y                     & bod & T   & F  & F  & T  & F  & F  & T  & T  & F  & T   & eod & eod \\ \hline\hline

X \olucidop{first} Y  &     & 1   & 1  & 1  & 1  & 1  & 1  & 1  & 1  & 1  & 1   &     &     \\ \hline
X \olucidop{last} Y   &     & 10  & 10 & 10 & 10 & 10 & 10 & 10 & 10 & 10 & 10  &     &     \\ \hline
X \olucidop{next} Y   &     &     & 2  & 3  & 4  & 5  & 6  & 7  & 8  & 9  & 10  & eod & eod \\ \hline
X \olucidop{prev} Y   &     & bod &    &    &    &    &    &    &    &    &     &     &     \\ \hline\hline

X \olucidop{fby} Y    &     & 1   & T  & F  & F  & T  & F  & F  & T  & T  & F   & T   & eod \\ \hline
X \olucidop{pby} Y    &     & T   & F  & F  & T  & F  & F  & T  & T  & F  & T   & 1   & eod \\ \hline\hline

X \olucidop{wvr} Y    &     & 1   &    &    & 4  &    &    & 7  & 8  &    & 10  &     &     \\ \hline
X \olucidop{rwvr} Y   &     & 10  &    &    & 8  &    &    & 7  & 4  &    & 1   &     &     \\ \hline
X \olucidop{nwvr} Y   &     &     & 2  & 3  &    & 5  & 6  &    &    & 9  &     &     &     \\ \hline
X \olucidop{nrwvr} Y  &     &     & 9  & 6  &    & 5  & 3  &    &    & 2  &     &     &     \\ \hline\hline

X \olucidop{asa} Y    &     & 1   & 1  & 1  & 1  & 1  & 1  & 1  & 1  & 1  & 1   &     &     \\ \hline
X \olucidop{nasa} Y   &     & 2   & 2  & 2  & 2  & 2  & 2  & 2  & 2  & 2  & 2   &     &     \\ \hline
X \olucidop{ala} Y    &     & 10  & 10 & 10 & 10 & 10 & 10 & 10 & 10 & 10 & 10  &     &     \\ \hline
X \olucidop{nala} Y   &     & 9   & 9  & 9  & 9  & 9  & 9  & 9  & 9  & 9  & 9   &     &     \\ \hline\hline

X \olucidop{upon} Y   &     & 1   & 2  & 2  & 2  & 3  & 3  & 3  & 4  & 5  & 5   & eod &     \\ \hline
X \olucidop{rupon} Y  &     & 10  & 9  & 9  & 8  & 7  & 7  & 7  & 6  & 6  & 6   & bod &     \\ \hline
X \olucidop{nupon} Y  &     & 1   & 1  & 2  & 3  & 3  & 4  & 5  & 5  & 5  & 6   & 6   & eod \\ \hline
X \olucidop{nrupon} Y &     & 10  & 10 & 9  & 9  & 9  & 8  & 7  & 7  & 6  & 5   & 5   & bod \\ \hline\hline

\olucidop{neg} X      &     & -1  & -2 & -3 & -4 & -5 & -6 & -7 & -8 & -9 & -10 & eod & eod \\ \hline
\olucidop{not} Y      &     & F   & T  & T  & F  & T  & T  & F  & F  & T  & F   & eod & eod \\ \hline
X \olucidop{and} Y    &     & 1   & 0  & 0  & 1  & 0  & 0  & 1  & 1  & 0  & 1   & eod & eod \\ \hline
X \olucidop{or} Y     &     & 1   & 2  & 3  & 5  & 5  & 6  & 7  & 9  & 9  & 11  & eod & eod \\ \hline
X \olucidop{xor} Y    &     & 0   & 2  & 3  & 5  & 5  & 6  & 6  & 9  & 9  & 11  & eod & eod \\ \hline
\end{tabular}
\caption{Example of Application of {\flucid} Operators to Bounded Streams}
\label{tab:flucid-op-examples}
\end{table}



\end{document}